\documentclass[12pt,a4paper]{article}
\pdfoutput=1
\usepackage{amsmath,amssymb,exscale}
\usepackage{amsfonts,amsthm}
\usepackage{graphicx, caption, float, epsfig}
\usepackage{mathrsfs}
\usepackage{slashed}
\usepackage{textcomp}
\usepackage[T1]{fontenc}
\usepackage[utf8]{inputenc}
\usepackage[table]{xcolor}
\usepackage[lowtilde]{url}
\usepackage{cancel}
\usepackage[font={small}]{caption}
\numberwithin{equation}{section}
\usepackage{verbatim} 
\usepackage{appendix}
\usepackage{hyperref}
\newcommand{\be}{\begin{equation}}
	\newcommand{\ee}{\end{equation}}
\newcommand{\ba}{\begin{eqnarray}}
	\newcommand{\ea}{\end{eqnarray}}
\newcommand{\bay}{\begin{array}{rcl}}
	\newcommand{\eay}{\end{array}}
\newcommand{\ra}{\rightarrow}

\begin{document}
	
\title{Stringy phenomenology with preon models}

\author{Risto Raitio \footnote{E-mail:risto.raitio@gmail.com}\\	
02230 Espoo, Finland}
	
\date{October 2, 2023}  \maketitle 
	
\abstract{\noindent
We compare, following Pati, global symmetries our topological supersymmetric preon model with the heterotic $E_8 \times E_8$ string theory. We include Pati's  supergravity based preon model in this work and compare the preon interactions of his model to ours. Based on preon-string symmetry comparison and preon phenomenological results we conclude that the fundamental particles are likely preons rather than standard model particles.}
	
\vskip 1.5cm
\noindent
	
\vskip 1.5cm
	
	
\newpage
	
%\tableofcontents
%\vskip 2.0cm
	
\section{Introduction}
\label{intro}
 
We analyze briefly in this note group structure of the heterotic $E_8 \times E_8$ string theory in 10D, and compare its symmetry structure to the composite models of quarks and leptons introduced by Pati \cite{Pati:1985} and us \cite{Rai10:2018, Rai10:2023a}. The former model proposes gauge interactions between preons, while we utilize topological concepts for preon interactions.

In spite of its immense experimental success the standard model (SM) has its  problems: it does not include gravity and supersymmetry (no sparticles found), it requires many arbitrary constants and parameters that are not derived from the theory itself, it does not explain the origin of lower scales such as $M_W$ and the origin of the number of generations, to mention some of them. String theory can in principle solve this kind of problems but a working phenomenology is missing.
 
Pati \cite{Pati:1985} and we \cite{Rai10:2023a} believe that the main cause of the SM difficulties is the choice of the fundamental fields as quarks and leptons. The quarks, the leptons and the Higgs particles should rather be interpreted as preon composites. As a bonus, we obtain to preliminary concordance between the preon models considered here and string theory.

We have introduced a topological Chern-Simons (CS) interaction for preon binding (our preons called from now on chernons) while the Pati model uses a non-abelian metacolor gauge force. CS model is preferred because we want to have background independent model, and secondly, the preonic phase of matter is assumed to occur at very high energy scale where no regular spacetime may not exist \cite{Rai10:2023f}. In addition, baryon asymmetry and the "hiddenness" of supersymmetry can be explained naturally \cite{Rai10:2023f}. 

We think that a topological approach to preon models is more far reaching than the $SU(4)_M$ in \cite{Pati:1985} with respect to string theory. The Calabi-Yau string manifold is topologically non-trivial. Topology affects the spectrum of particles and determines whether the theory contains massless particles, like scalars and the graviton. Our use of a topological model is, at this stage, far from string theory as such, but it may be a first step, among others, to stringy particle phenomenology with applications to cosmology.

This note is organized as follows. The section \ref{phenom} is divided in three subsections. We first take a glimpse in heterotic $E_8 \times E_8$ string group theory. In subsection \ref{patimodel} we recap the Pati preon model. In the next subsection \ref{chernonmodel} we present a summary of the features of our preon model. The differences in the preon models considered here are pointed out. Conclusions are given in section \ref{conclusions}.

%****************************************
\section{Strings and preon phenomenology}
\label{phenom}

We begin with a concise recap of heterotic string theory $E_8 \times E_8$ in 10D. This provides a yardstick for finding similar structures between strings and preons/chernons.

\subsection{Very brief string recap}
\label{strbgr}

We start from the heterotic string theory $E_8 \times E_8$ in D = 10. It can be compactified to $M^4 \times K$ where $M^4$ is the four-dimensional Minkowski space and $K$ a compact six-dimensional Calabi-Yau manifold with SU(3) holonomy \cite{Witten:1985}. This compactification leaves an unbroken N = 1 local supersymmetry at the Planck, or compactification, scale. Further, it breaks $E_8 \times E_8$ into $E_8 \times E_6$ if $K$ is simply connected, and into a lower symmetry such as $E_8 \times SU(3) \times SU(2)_{\rm{L}} \times [U(1)]^3$ or $E_8 \times SU(3) \times SU(2)_{\rm{L}} \times [U(1)]^2$, if $K$ is multiply connected. The topology of $K$ determines the massless zero mode matter superfields in 4D. These are in the form of several copies of 27's and $\overline{27}$'s of $E_6$ (even when it is broken). The number of generations is $N = n_{27} - n_{\overline{27}}$. This is given by half the Euler characteristic of $K$ which is low enough if $K$ is multiply connected. Models with 1, 2 and 4 generations are known  \cite{Witten:1985}. Manifolds giving rise to three generations have been constructed in \cite{Strominger:1985}.

The anomaly-free $E_8 \times E_8$ string theory leads in 4D to chiral fermions (i.e. N > 0), which are in sets of 27 of $E_6$. These include the known quarks and leptons belonging to 16 of $SO(10) \in E_6$ plus some exotic fermions. The higher dimensional string theories can determine uniquely the number of families N, if one can solve the dynamics of compactification on the basis of the topology of $K$. 

\subsection{The Pati preon model}
\label{patimodel}

An economical preon model which is based on N = 1 supergravity in 4D and which incorporates the two-scale idea is proposed in \cite{Pati:1984}. The model introduces four left- plus four right-handed chiral superfields, each transforming as a fundamental or some fixed representation $r$ of a non-abelian metacolor gauge symmetry $G_M$. The scale parameter $\Lambda_M$ of $G_M$ is determined to be rather high, about $10^{12-16}$ GeV, from renormalization group equations. The familiar flavor $W_{\rm{L}, \rm{R}}$, color, and hypercolor particles, associated with an effective SU(4) gauge symmetry, are interpreted as preonic composites of very small sizes $\sim 1/\Lambda_M$, bound by the primordial metacolor gauge force. Due to a two-fold replication, the model predicts four quark-lepton families ($e, \mu, \tau$ and $\tau '$),  i.e. effectively eight flavors. Barring mixing, the $e$ and the $\mu$ families have very small sizes $\sim 1/\Lambda_M$, while the $\tau$ and the $\tau '$ families have large sizes $\sim 1/\Lambda_H \sim $1/TeV. The model possesses a mechanism for interfamily mass hierarchy. 

Pati has shown that the precise field content of four left- plus four right-handed superfields of the model sketched above can be derived from the 10D, $E_8 \times E_8$ string theory provided that the compactification to 4D leads to two copies of 27's of $E_6$ and that $E_6$ breaks at the compactification scale to a subgroup $G_0 = SU(4)_M \times \tilde{G}$ where $\tilde{G}$ is either $[U(1)]^3$ or $SU(2) \times [U(I)]^2$, or $[U(1)]^2$ or $SU(2) \times U(1)$. The symmetry $\tilde{G}$, on the other hand, breaks completely at $\Lambda_M$, dynamically due to preon condensates. 

Further details of derivation of this preon model emerging from the $E_8 \times E_8$ string theory are given in \cite{Pati:1984}.

\subsection{The chernon model}
\label{chernonmodel}

The chernon model was developed originally from bottom up so that three chermons of certain charge (namely $\pm 1/3$) can produce the quark charges 2/3 and -1/3. Later we found that the model resembles closely the global supersymmetric Wess-Zumino model \cite{WZ:1974}, which can be extended to supergravity.\footnote{The model is a first generation case. A three generation version including dark matter is under consideration.}

The next phase in developing the model was to make it functioning in early cosmology. Introducing a binding energy scale near reheating $\Lambda_{cr} \geq 10^{10-16}$ GeV for the model, T-duality was introduced \cite{Rai10:2022}. Thirdly, Above $\Lambda_{cr}$ the universe was in a form having topological interactions. Any metric needs not necessarily be defined. Therefore, 2+1 dimensional Chern-Simons interactions between the chernons, inside a 3+1 dimensional spacetime, were introduced. Chernons with certain spontaneous symmetry breaking turned out to give a natural explanation for baryon asymmetry \cite{Rai10:2023a}. Furthermore, the chernon model explains why supersymmetry has not been observed in nature \cite{Rai10:2023f}. The preon interaction is the main difference between the Pati model(s) and ours. Nevertheless, both models have similar numerical behavior in UV and IR. 

Chern-Simons-Maxwell (CSM) models have been studied in condensed matter physics (for original references see \cite{Rai10:2023a}). In this note we apply the CSM model in particle physics phenomenology in the early universe. The following CS-QED action has been defined

\begin{align}
	{S}_{{\rm {CS-QED}}}^{{\rm SSB}} & =\int d^{3}x\biggl\{-\frac{1}{4}F^{\mu \nu}
	F_{\mu \nu }+\frac{1}{2}M_{A}^{2}A^{\mu }A_{\mu } \nonumber \\
	&~~ -\frac{1}{2\xi }(\partial^{\mu }A_{\mu })^{2}+\overline{\psi }_+(i\cancel\partial -m_{eff})\psi _{+} \nonumber \\
	&~~ +\overline{\psi }_{-}(i\cancel\partial +m_{eff})\psi _{-}+ \frac{1}{2}
	\theta \epsilon ^{\mu v\alpha }A_{\mu }\partial _{v}A_{\alpha }  \nonumber \\
	&~~ +\partial ^{\mu }H\partial _{\mu }H-M_{H}^{2}H^{2} +\partial ^{\mu }\theta
	\partial _{\mu }\theta -M_{\theta }^{2}\theta ^{2} \nonumber \\
	&~~ -2yv(\overline{\psi }_{+}\psi _{+}-\overline{\psi }_{-}\psi _{-})H-e_{3}\left( \overline{\psi }
	_{+}\cancel A\psi _{+}+\overline{\psi }_{-}\cancel A\psi _{-}\right) \biggr\}  \label{actionMCS3}
\end{align}
where the mass parameters 
\begin{equation}
	M_{A}^{2}=2v^{2}e_{3}^{2},~~m_{eff}=m_{ch}+yv^{2},~~M_H^{2}=2v^{2}(\zeta +2\lambda v^{2}),~~M_{\theta }^{2}=\xi M_{A}^{2}
	\label{masspar}
\end{equation}
depend on the SSB mechanism: $v^{2}$, the vacuum expectation value, and $y$ is the parameter that measures the coupling between fermions and Higgs scalar. Being a free parameter, $v^{2}$ indicates the energy scale of the spontaneous breakdown of the $U(1)$ local symmetry. The Proca mass $M_A^2$ represents the mass acquired by the photon through the Higgs mechanism. The Higgs mass, $M_{H}^{2}$, is associated with the real scalar field. The Higgs mechanism also contributes to the chernon mass $m_{ch}$, resulting in an effective mass $m_{eff}$. There are two photon mass-terms in (\ref{actionMCS3}), the Proca and the topological one. Now the idea of (\ref{actionMCS3}) is that the parameters of the model can be chosen such that the photon mass is high enough to create an attractive, very short range Yukawa force which is stronger than the repulsive Coulomb force between equal charge chernons. 

The chernon-chernon scattering amplitude in the non-relativistic approximation is obtained by calculating the t-channel exchange diagrams of the Higgs scalar and the massive gauge field. The propagators of the two exchanged particles and the vertex factors are calculated from the action (\ref{actionMCS3}). The gauge invariant effective potential for the scattering considered is obtained in \cite{Kogan, Dobroliubov}
\begin{equation}
	V_{{\rm MCS}}(r)=\frac{e^{2}}{2\pi }\left[ 1-\frac{\theta }{m_{ch}}\right]
	K_{0}(\theta r)+\frac{1}{m_{ch}r^{2}}\left\{ l-\frac{e^{2}}{2\pi \theta }%
	[1-\theta rK_{1}(\theta r)]\right\} ^{2} 
	\label{Vmcs}
\end{equation}
where $K_{0}(x)$ and $K_{1}(x)$ are the modified Bessel functions and $l$ is the angular momentum ($l=0$ in this note).  

One sees from (\ref{Vmcs}) the first term may be positive or negative while the second term is always positive. The function $K_{0}(x)$ diverges as $x \ra 0$ and approaches zero for $x \ra \infty$ and $K_{1}(x)$ has qualitatively similar behavior. For our scenario we need negative potential between equal charge chernons. We can achieve this if we require the condition
\be
\theta \gg m_{ch}
\label{condition}
\ee
The attractive equal charge potential plays the key role when chernons begin to form quarks and leptons \cite{Rai10:2023a}. 

For applications to condensed matter physics, one must require $\theta \ll m_{e}$, and the scattering potential given by (\ref{Vmcs}) then comes out positive \cite{Beli_D_F_H}.

\begin{figure}
	\centering
	\captionsetup{width=.8\linewidth}
	\includegraphics[width=10cm]{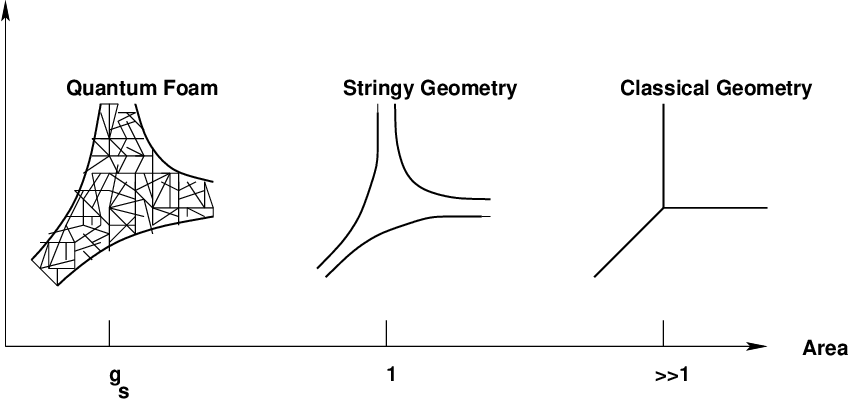}
	\caption{\small The diagram shows how the geometry may vary as one changes the area scale, drawn in string units. Our model is constructed for area scale > 1. Figure is from \cite{Vafa:2008}.}
	\label{fig:figure1}
\end{figure}

%The arguments of subsection \ref{strbgr} apply in this subsection as well as in the Pati model case in subsection \ref{patimodel}. Instead, 
Neither the preon nor the chernon binding interactions are fully studied. We discuss briefly our case. The global attractive potential (\ref{Vmcs}) at tree level is obtained after the introduction of the Higgs mechanism in the context of the MCS-QED \cite{Beli_D_F_H}. This should allow approximate Yukawa bound states of chernons below the scale $\Lambda_{cr}$ and free chernons above the scale $\Lambda_{cr}$. The eigenvalues of a Yukawa potential in Schr\"{o}dinger equation are $E = -\frac{1}{2} \mu g^2 \epsilon_{n,l}$ with
\be
\epsilon_{n,l} = -\frac{1}{n^2} + 2 \delta - \frac{1}{2}\big(3n^2 - l(l+1)\big)\delta^2
\label{enerrgye}
\ee
where $\mu$ is the reduced mass of the particle, $\delta = a_0/D$, $a_0$ is the Bohr radius of the bound state, $D$ the range of the potential and $g$ the strength of the force \cite{Naps:2021}. The bound state mass would be of the order of chernon mass. The energy scale of the composite formation is, however, $\Lambda_{cr}$. This may be the point where smooth continuum theory changes, see figure \ref{fig:figure1}. We should seek some more general spacetime framework \cite{Vafa:2008}. This question is beyond the scope of this note.

\section{Conclusions and outlook}
\label{conclusions}

In this note we have sought connection between strings and chernons. We compare first the two models of Pati and us discussed in subsections \ref{patimodel} and \ref{chernonmodel}, respectively. In Pati's model, the existence of vacuum solutions $M^4 \times K$ at the tree level are known. Such a compactification leaves an unbroken N = 1 local supersymmetry at the compactification scale. It breaks $E_8 \times E_8$ into $E_8 \times E_6$ if K is simply connected and $E_6$ breaks at the compactification scale to a subgroup $G_0 = SU(4)_M \times \tilde{G}$. Pati identifies the symmetry group $SU(4)_M$ with the metacolor gauge symmetry. The field theory with this gauge group is asymptotically free in UV and confining in IR \cite{Ayyar:2017}. The spectrum of states of $SU(4)_M$ may be too rich for a preon model. Numerically, our model has similar UV/IR behavior as the Pati model.\footnote{Numerical agreement may or may not indicate real connection between models. For inflation, the Starobinsky model agrees well with the supergravity model \cite{Rai10:2023a}.}

The global symmetry analysis of our model with one fermion generations is weaker than Pati's. Our model can be extended to include supergravity but we have chose the binding interaction from Chern-Simons model. The number of generations depends on the compactification of the topology of $K$. We have done standard quantum mechanics for approximate binding at scale $m_{ch}$. At scales above  $\Lambda_{cr}$ we believe the topology spacetime manifold changes, presumably towards string topology. On the other hand, we have found the following experimental support for the chernon model: (i) it reduces to the standard model at accelerator energies \cite{Rai10:2018}, (ii) it has proposal for the dark sector (not observed/understood yet) \cite{Rai10:2023a}, (iii) it explains the baryon asymmetry in the universe in a natural way \cite{Rai10:2023a}, (iv) it provides T-dual topological evolution of the universe \cite{Rai10:2022}, and (v) it explains why supersymmetry has not been found in experiments \cite{Rai10:2023f}.

These agreements are found using phenomenological methods rather than by mathematically rigorous proofs. It is our understanding that supersymmetry and string theory have some direct support from experiments. 

A crucial question for our model is can Chern-Simons theory be derived from the topology of $K$ in some limit. If so this may provide a way to study aspects of string theory using Chern-Simons theory as a tool. This has been studied  in \cite{Grassi:2004, Witten:2003}. \footnote{We mention to the curious reader that CS theory and the Kodama state have been investigated for quantum gravity and in connection of the wave function of the universe \cite{Witten:2003a, Alex&al:2012}.}

\end{document}